\documentstyle[epsfig]{mn}

\title[Time-resolved Spectroscopy of the X-ray Binary AC211]{Time-resolved Spectroscopy of the M15 X-ray Binary AC211/X2127+119}
\author[M. A. P. Torres et al.]{M. A. P. Torres$^{1}$, P. J. Callanan$^{1}$, M. R. Garcia$^{2}$ \\
$^{1}$Physics Department, University College, Cork, Ireland\\
$^{2}$Harvard-Smithsonian Center for Astrophysics, 60 Garden St, Cambridge, MA 02138}

\begin{document}

\maketitle

\begin{abstract}
We present time-resolved spectroscopy acquired during two epochs
(spaced apart by $\sim15$ days) of the eclipsing Low Mass X-ray Binary
AC211/X2127+119 in the globular cluster M15. The spectra show
variations in the He{\sc ii}~$\lambda4686$~emission line not only
modulated on the orbital period, but also on time-scales of a few
days. During the first epoch of observation, the emission line
shows a strong S-wave superimposed on the average double-peaked
profile. The line exhibits no evidence of rotational disturbance at the orbital phases when the eclipse is observed in the optical
continuum. During the second epoch, no double-peak or S-wave
component is present. The He{\sc i} absorption lines detected by other
authors are not present in our spectra. A Doppler image of the He{\sc
ii}~$\lambda4686$ for the first epoch supports the presence of the
accretion disc. No hotspot is detected, although enhanced emission at
${V_X}=30$~km s$^{-1}$, ${V_Y}=160$~km s$^{-1}$ is observed. We
discuss the implications of this emission in the context of an X-ray
heated donor star, in which case a high mass ratio and neutron star
primary are implied. Finally, we speculate on the possibility of a
misaligned secondary star in AC211.

\end{abstract}
 

\section{Introduction}

Low Mass X-ray Binaries (LMXBs) are systems which harbour an accretion
disc through which mass from a low mass secondary
(M${_2}\la1~$M$_\odot$) is accreted onto a neutron star or black hole
primary. It is commonly accepted that LMXBs in the galactic disc and
halo are the end product of binary evolution (see the review by
Verbunt \& van den Heuvel 1995). From early X-ray studies it is known
that the rate of occurrence of persistently bright
(L${_X}>10^{36}$~erg s$^{-1}$) LMXBs in globular clusters is $\sim100$
times higher than in other regions of our Galaxy (Katz 1975; Clark
1975). Therefore a different mechanism which enhances LMXB formation
must operate in globular clusters. According to binary evolutionary
models, it is currently believed that the high stellar densities and
enhanced rates of star interaction in globular clusters provide a
favourable environment for the formation of LMXBs via tidal capture
through close encounters between neutron stars and ordinary stars
(Fabian, Pringle \& Rees 1975). So far, 13 persistently bright LMXBs
have been discovered in globular clusters. Twelve of them, with a
collection of their properties, are listed in tables 1 of Grindlay
(1993) and Deutsch et al. (2000). The 13$^{th}$ object is the recently
discovered M15 X-2 (White \& Angelini 2001). Most of the globular
cluster LMXBs are type-I X-ray burst sources, indicating that the
compact object is an accreting neutron star. The increased effort to
identify their optical counterparts is a critical part in the process
of obtaining the dynamical properties of these sources (see
e.g. Deutsch et al. 2000). Furthermore, their study provides important
information about the evolution of the host globular cluster since
close binaries can have a dramatic effect on the cluster dynamics
(Elson et al. 1987; Hut et al. 1992).

X2127+119 was one of the first X-ray sources identified in a globular
cluster (Giacconi et al. 1974) and the first ever globular cluster
LMXB optically identified (Auri\`ere, Le F\'evre \& Terzan 1984;
Auri\`ere et al. 1986, Charles, Jones \& Naylor 1986). Located within
2 arcsec of the crowded core of M15 (NGC 7078), the optical
counterpart, AC211, is a ${\it V}\sim15$ magnitude variable star. This
makes AC211 one of the optically brightest LMXBs. Soon after its
discovery, a periodicity of 8.5 hr was observed in the optical
(Ilovaisky et al. 1987, Naylor et al. 1987) and X-rays (Hertz 1987),
and initially associated with the orbital period of the
system. However, Ilovaisky et al. (1993) showed through further
optical photometry that the true orbital period of AC211 is 17.1
hr. The best ephemeris to date is provided by Homer \& Charles
(1998). The optical light curve of AC211 exhibits the largest amplitude
orbital modulation of any X-ray binary (1.8 mag in {\it U}), showing a
symmetrical partial primary eclipse and a broad dipping structure
centred at phase 0.5 (Ilovaisky et al. 1993; Homer \& Charles
1998). AC211 has been also studied in the UV (see e.g. Naylor et
al. 1988, 1992) and EUV (Callanan, Drake \& Fruscione 1999).

The low X-ray to optical luminosity ratio ($L{_X}/L{_{opt}}$$\sim20$)
suggests that AC211 is an Accretion Disc Corona (ADC) source (see
White \& Holt 1982 for an outline of this model). The eclipse and
dipping components of the light curves together with the orbital
modulation of the X-ray absorption column density (Hert \& Grindlay
1983, Ioannou et al. 2002a) can be easily explained if the inclination
of AC211 is high. Then the secondary star and the outer asymmetric rim
of the disc will partially obscure the ADC X-ray source causing the
eclipse and dips, respectively. However, the detection by {\it Ginga}
of an X-ray burst (with radius expansion) associated with AC211
implied that the compact object was a directly visible neutron star
(Dotani et al. 1990; van Paradijs et al. 1990), a result apparently
inconsistent with the ADC model. Recently, White \& Angelini (2001),
through observations with the {\it Chandra} X-ray Observatory have
resolved X2127+119 into two X-ray sources: the LMXB associated with
AC211 and a new persistent LMXB, M15 X-2. Both of these are separated
by 2.7 arcsec. Since M15 X-2 is 1.6 times brighter in X-rays than
AC211, its X-ray flux significantly contributes to the flux of the
unresolved sources. This artificially increases the X-ray luminosity
of AC211, contaminates the X-ray spectrum and reduces the amplitude of
the orbital modulation in the X-ray light curve obtained in previous
observations. Furthermore, since M15 X-2 shows no clear modulation in
its X-ray light curve, it is reasonable to assume that it is a
low-inclination LMXB and we can see the compact object directly. Thus
the bright X-ray bursts observed by {\it Ginga} (Dotani et al. 1990)
and {\it RXTE} (Smale 2001) can be associated with a neutron star in
M15 X-2. These arguments provide a natural explanation for some of the
AC211's unusual multi-wavelength properties and further strengthen the
ADC scenario.

However, despite these extensive studies, many of the fundamental
physical parameters of AC211 remain unknown. This is largely due to
the fact that in AC211, as in most of the persistently bright LMXBs,
the optical flux generated in the disc by X-ray reprocessing
overwhelms the absorption line spectrum of the secondary star. This
makes it difficult to measure the radial velocity curve of the
secondary. The background contamination from the surrounding field
stars in M15 represents an annoying extra problem. Optical
spectroscopy of AC211 by Naylor et al. (1988) showed the presence of
emission at He{\sc ii}~$\lambda4686$, Balmer lines (overpowered by the
background) and He{\sc i} absorption features. From an analysis of the
He{\sc i}~absorption features and assuming an origin in the outer
accretion disc, they claimed a systemic velocity for AC211
blue-shifted by $-150\pm10$~km s$^{-1}$ with respect to the globular
cluster. Such a high $\gamma$ velocity suggested tidal interaction as
the mechanism of formation of the binary system accompanied by
subsequent cluster ejection. However, observations by Ilovaisky et
al. (1989) clearly displayed variations in the velocity of the He{\sc
i} feature which showed that it was not representative of the true
AC211 systemic velocity. Two alternative explanations for the He{\sc
i}~$\lambda4471$ velocity have been suggested: Fabian, Guilbert \&
Callanan (1987) pointed to outflowing mass above the accretion disc as
cause of the He{\sc i}~absorption. In favour of this hypothesis was
the discovery from {\it HST} observations of P-Cygni profiles in the
H$\beta$ and C{\sc iv}~$\lambda1550$ emission lines (Downes, Anderson
\& Margon 1996; Ioannou et al. 2002b). On the other hand, Bailyn,
Garcia \& Grindlay (1989) argued that the blueshifted He{\sc i} can
originate in an outflowing gas from the $L_2$ Lagrangian point owing
to a common envelope in the binary system. However, recent re-analysis
of previously published optical spectroscopy by van Zyl et al. (2002)
shows an orbital modulation for the He{\sc i}~$\lambda4471$
inconsistent with this model (see also Callanan, Naylor \& Charles
1990).

Furthermore, no definitive picture of the accretion disc structure for
AC211 has emerged. So far, for example, the origin of the dipping
behaviour observed at phase $\sim 0.5$ is unknown: it may be due to a
bulge caused by the re-impact of a gas stream which overflows the
initial impact with the accretion disc. The lack of reliable dynamical
properties for this system and observational details of the disc
structure motivated us to carry out phase-resolved spectroscopy of the
He{\sc ii}~$\lambda4686$ feature and the He{\sc i}~absorption
lines. We proceed by giving in Section 2 a detailed discussion of the
observations and data reduction. In Section 3 the analysis of the
spectra is presented. A discussion of our results and a possible model
that could account for some of the properties of AC211 is given in
Section 4. Finally, a summary of our conclusions is presented in
Section 5.

\section{Observations and Data Reduction}

AC211 was observed with the 4.5-m Multiple Mirror Telescope (MMT) on
Mount Hopkins in two runs during 1996 August 10-14 UT and August 31,
September 1-2 UT. Despite the variable observing conditions and
mediocre seeing throughout all the observing nights (with a seeing
FWHM of about 2 arcsec), the nights were good enough for spectroscopic
observations. For both runs, we used the Blue Channel Spectrograph
equipped with the Loral 3072x1024 CCD and the 832 grooves mm$^{-1}$
grating in second order. A CuSO$_4$ filter was used to block the first
order light. This setting provided a spectral coverage of 1000 \AA, a
dispersion scale of 0.35 \AA~pix$^{-1}$ and a spectral resolution of 1
\AA~FWHM. The slit was rotated at position angles which reduced
contamination from nearby cluster stars and the spatial binning of the
chip was carried out as a function of the seeing conditions. The
exposure times were typically 600 and 900 seconds. HeNeAr arc lamp
spectra were taken regularly through the nights in order to determine
the wavelength calibration. A record of the observations and
instrumental setup is provided in Table 1.

The images were bias and flat-field corrected with standard {\sc
iraf}\footnote{{\sc iraf} is distributed by the National Optical
Astronomy Observatories.} routines. Spectra were extracted with the
{\sc iraf kpnoslit} package. The extraction of AC211 spectra and four
nearby field stars was a process necessarily interactive and it was
repeated several times image by image until an optimum result was
obtained. The apertures were edited for each star and sized to
maximize the number of pixels contributing to the object. Next, the
spectra profiles were traced and the spectra were extracted without
background subtraction because the slit did not include any sky
region. Note that we expect a negligible sky contribution since the
spectral interval observed is free of strong line bands (see
e.g. Massey \& Foltz 2000) and the main source of background is the
unresolved globular cluster light and that from AC207 (Auri\'ere \&
Cordoni 1981): the AC211 spectra presented here suffer considerably
from contamination by the latter.

The pixel-to-wavelength calibration was derived from cubic spline fits
to about 19 arc lines, giving a root-mean square deviation of typically
$\la0.03$~\AA. The wavelength scales of neighbouring arc spectra were
interpolated in time and the accuracy of the whole process was checked
by cross-correlation of the normalized sum of two field star spectra
in the range $\lambda\lambda4900-5000$ with the individual spectra of
the corresponding star. The systematic velocity errors obtained
from the scatter of the measured zero velocities were less than 3 and
5 km s$^{-1}$ for the first and second run, respectively. Two spectra
were rejected because they gave residual velocities of 7 and
4-$\sigma$.

We also found that in a few cases the trace was lost when extracting the
spectra. In general, this happened near the ends of the slit where the
throughput of the spectrograph  drops. We removed these regions in the
spectra by masking them. On the night of September 1, we suffered
difficulties because of condensation inside the CCD control box which
introduced spurious negative features in the object and arc frame
spectra. Because of that, we rejected this night from the
analysis. Our final data set consists of 40 and 19 spectra of AC211
for the first and second run respectively.

For the subsequent spectra analyses we used the MOLLY package. The
spectra were rebinned into a constant velocity scale and normalized by
dividing them by the result of fitting a low order spline to the
continuum after masking the Balmer absorption lines and the He{\sc
ii}~$\lambda4686$ emission line in the case of AC211. To fold the
spectra on the  orbital period, we used the linear ephemerides from
Homer et al. (1998; 2002) to determine the orbital
phases:
$$\phi=0.0,~JD=2447790.964(2) \pm n \times 0.7130207(5)~d$$
where $\phi=0.0$ corresponds to closest approach of the secondary to the
observer.

\section{Results}

\subsection{The mean spectrum}
A comparison between the mean spectrum of AC211 over the two runs is
shown in Fig. 1. The only feature in the spectrum with unquestionable
origin from AC211 is the He{\sc ii}~$\lambda4686$ emission line. For
the first run, this line shows a full width zero intensity (FWZI) of
865~km s$^{-1}$, a FWHM=516~km s$^{-1}$ and a mean equivalent width
(EW) of 1.0 $\pm$ 0.3~\AA, with the error accounting (from now on) for
line variability. The mean line  shows a double-peaked profile with a
peak-to-peak separation of $288 \pm 6$ km s$^{-1}$ and enhanced
emission on the blue-shifted peak.  For the second run, we measured a
FWZI=800~km s$^{-1}$, a FWHM=373~km s$^{-1}$ and an EW=0.6$\pm$
0.1~\AA~for a line which shows no double peak profile. Note that the
EW estimates in particular are underestimated because of contamination
from AC207.

The spectral range observed was chosen to cover the He{\sc i}
absorption lines reported at $\lambda\lambda4388,4471,4713,4921$ by
Naylor et al. (1988, 1989) and Ilovaisky (1989). A meticulous search
for them in the mean and individual spectra was unsuccessful. An
absorption feature at $\lambda4469$ (rest frame wavelength) was
discerned in the AC211 spectrum, but it is also present in the field
stars (see Fig. 2), indicating that the absorption in AC211 is due to
contamination from the globular cluster background. We suggest that
this feature is (partial or totally) due to Ti{\sc
ii}~$\lambda4468.5$. Titanium lines are detected in the red giant
members of M15 (see e.g. Cohen 1979; Sneden et al. 2000) and its
abundance is even enhanced for Blue Horizontal Branch stars (Behr,
Cohen \& McCarthy 2000). We measure an equivalent width of 0.18 $\pm$
0.05~\AA~for the AC211 spectra and 0.19 $\pm$ 0.03, 0.27 $\pm$ 0.06,
0.24 $\pm$ 0.03 and 0.12 $\pm$ 0.02~\AA~for the field stars (see Auri\'ere \&
Cordoni 1981) AC160, AC11, AC623 and AC256 respectively.

Most of the absorption profile for the prominent H$\beta$ line is also
due to the background contamination, but we will show below that part
of this line has its origin in the binary system.

\subsection{Phase dependent and long-term profile variations}

In order to examine the variability of the line profiles with orbital
phase, we folded the He{\sc ii}~$\lambda4686$ and the H$\beta$
profiles into 15 orbital phase bins by using the ephemeris of Sec
2. Fig. 3 displays the results obtained for each run in the form of
trailed spectrograms where two obvious features appear. In the
upper-left panel, an S-wave is present in the He{\sc ii}~$\lambda4686$
profiles. This line component reaches maximum blueshift around phase
0.75 and maximum redshift around 0.25, with intensity increasing
relative to the continuum flux between orbital phases $0.68-0.94$. The
modulation in intensity explains the asymmetry in intensity of the red
and blue components of the double-peaked emission for the first run
when averaged over the orbit. Another interesting characteristic in
the behaviour of the He{\sc ii}~$\lambda4686$ profile is the absence
of clear signs of rotational disturbance during the eclipse of
the accretion disc by the donor star (which ranges in orbital phases
from $0.8-0.15$; Ilovaisky et al. 1993). The lack of a disturbance in
the line implies that at least some of the emission is formed mainly
in non-eclipsed vertical regions on or over the accretion disc.

The radial velocity curve and EWs of the He{\sc ii}~$\lambda4686$ are
plotted in Fig. 4. The radial velocities were determined by
cross-correlation of each individual profile with a Gaussian template
having the same width as the mean line for the corresponding run. In
the case of the first run, these  velocities are modulated clearly
with phase, ranging from a minimum of -197~km s$^{-1}$ to a maximum of
$-6$~km s$^{-1}$. It is the S-wave, which moves and varies in strength
across the profile, that is the cause of these variations, while
the main body of the line shows no clear modulation.

The plot of the EW versus orbital phase shows that the EW is larger
during the eclipse of the accretion disc by the secondary star, with
higher EWs centered around phase 0.9 and lower EWs observed at phase
$\sim0.3$. This behaviour suggests again that a considerable fraction
of the He{\sc ii} emitting region is not closely confined to the
orbital plane: an intrinsically variable He{\sc ii} line flux combined
with an eclipse by the secondary of the continuum from the accretion
disc could explain the observed modulation in the EW. Nevertheless,
the possibility of apparent EW variability introduced by seeing
variations may also affect these measurements.

Despite the fact that a large fraction of the H$\beta$ line is due to
the cluster background, the folded spectra in the upper-right panel of
Fig. 3 show evidence for orbital motion and, therefore, for a
contribution from AC211. In particular, this line develops a blue
absorption wing around orbital phase 0.25. This is in contrast with
the results of van Zyl et al. (2002) who found no clear dependence of the
H$\beta$ line on the orbital cycle.

Because of poor orbital phase coverage, the observations during the
second run (lower panels of Fig. 3) are of less use in trying to
measure orbital or longer term variations in the line profiles. In
this regard, the trailed spectra do not sample the S-wave component at
He{\sc ii}~$\lambda4686$ or the blue absorption wing at H$\beta$
adequately. Fortunately some information can still be  extracted since
the observations during both runs match at four phase intervals
($0.0-0.07$, $0.13-0.2$, $0.4-0.47$ and $0.53-0.67$). Comparison of
the He{\sc ii} line profiles at the same phase intervals showed no
recurrence of the profile shape. This result points to changes in the
gas dynamics of the binary system over a time-scale of order a few
days. Since the Rossi X-ray Timing Explorer All-Sky Monitor light
curve of X2127+119 shows no significant variation in the X-ray
flux/hardness ratio between both epochs of observation, we think it
improbable that the differences observed above between the line
profiles are due to changes in the degree of X-ray irradiation.

\subsection{Doppler tomography}

In order to gain insight into the He{\sc ii}~$\lambda4686$ emitting
regions in AC211, a tomogram of this line was constructed by using the
Doppler tomography technique (Marsh and Horne 1988). We applied the
maximum-entropy method of building the tomograms from the profiles
obtained during the first run. Note that this technique was not used
with the profiles from the second run due to the poorer orbital coverage
($<50\%$). Also we did not attempt to combine the data (in order to improve
the orbital coverage) since there is evidence of long-term
variations in the line profiles (Sec 3.2).

To obtain a reliable brightness distribution, the phase-resolved
spectra acquired during the eclipse of the accretion disc by the
secondary star should not be included (see e.g Marsh \& Horne
1990). However, because the eclipse appears to have no effect
on the velocity profiles of the He{\sc ii}~$\lambda4686$ line, we
decided to build the map using the whole data set of 40
spectra. During the process of Doppler tomography, the $\gamma$
velocity was adjusted to the cluster mean velocity (-107.8 km
s$^{-1}$, Gebhardt et al. 1997). This value is in agreement with the
velocities obtained by fitting a Gaussian to the average profile. We
note that such a velocity for the He{\sc ii} line, in agreement with
the cluster velocity, argues against a high systemic velocity for
AC211. Fig. 5 displays the resultant tomogram.  The map shows a faint
ring of emission (a signature of emission arising from a rotating
accretion disc) and a region of enhanced intensity located at
approximately ${V_X}=30$~km s$^{-1}$, ${V_Y}=160$~km s$^{-1}$. This
tomogram should be considered an approximate reconstruction of the
velocity dependent line brightness distribution, since it does not
strictly satisfy the assumptions that underlie the Doppler tomography
technique, for example, that all motion is confined to the
orbital plane or that all points in the system are equally visible at
all times. In this regard, the image could also be affected by the
inclusion of data observed during orbital phases $0.35-0.65$, when the
accretion disc may partially mask the secondary star (Ilovaisky et
al. 1993).

\section{Discussion}

Our He{\sc ii}~$\lambda4686$ EWs are lower than the 1.7 \AA~reported
by Bailyn, Garcia \& Grindlay (1989) and considerably lower than the
9.7~\AA~measured by Downes et al. (1996) using {\it HST}
observations. Since our observations were taken with inferior seeing
than those refered to above, our spectra may be affected by higher
background contamination which artificially increases the
continuum. We attempted to decontaminate the MMT spectra of AC211 (out
of eclipse) by subtracting different amounts of local background in an
attempt to match the resulting continuum to the flux distribution of a
black body of T$_{eff }$=15,000 K, found by Downes et al. (1996). This
resulted in He{\sc ii}~$\lambda4686$ EWs increasing only by a factor
of two, still at least a factor of 5 lower than the values measured
from the {\it HST} observations (note that the MMT spectrum we
selected for comparison and the {\it HST} spectrum were both obtained
at similar orbital phases). This result suggests that the differences
observed in the measured EWs are intrinsic to AC211 and reflect
long-term dramatic changes in the emitting regions. This may be
related to the changes observed between the two runs in the overall
averaged line profile and points to significant changes in the form of
the gas flow in the system. Also, in comparison to previous
observations, no He{\sc i} absorption lines are detected in the
spectra (even after our attempts at decontamination), suggesting again
long-term variability in the gas regions where these lines are
formed. However, because of the lack of simultaneous X-ray
observations, we cannot rule out the possibility that the differences
in strength of the He{\sc ii}~$\lambda4686$ line and the
presence/absence of the He{\sc i} absorption lines may be caused
partially or totally by changes in the flux or hardness of the X-ray
flux (in contrast to the discussion in Sec 3.2).

Keeping in mind the caveats mentioned in the previous section
concerning the Doppler image, we now discuss the possible origins of
the He{\sc ii}~$\lambda4686$. In contrast to the tomograms presented
by van Zyl et al. (2002), it is clear that flux from the  accretion
disc contributes to this line. The tomogram (Fig. 5) shows a diffuse
ring with low velocities ($\sim200$~km s$^{-1}$). This is consistent
with emission from the outer parts of a high inclination accretion
disc as expected from the observed eclipses in the optical light
curve. A more striking question is the origin of the enhanced emission
observed at ${V_X}=30$~km s$^{-1}$, ${V_Y}=160$~km s$^{-1}$.  This
bright region produces the S-wave component in the trailed spectra of
the He{\sc ii} line (Fig. 4), which reaches maximum velocity near
phase 0.75 and minimum near phase 0.25.  In this regard, our S-wave
corresponds to the so-called component B observed by van Zyl et
al. (2002) in their He{\sc ii}~$\lambda4686$ trailed spectra. Since
the orbital modulation of the radial velocities is that expected for
emission arising from the X-ray heated face of the secondary star, we
decided to overplot on the Doppler image different outlines of the
Roche lobe of the secondary star in order to constrain the physical
parameters of the system.  Knowing the orbital period, the position
and the outline of the Roche lobe were calculated firstly by varying
the inclination angle of the system between 60 and 90 degrees and
assuming that the primary star is a neutron star of mass
${M_1}=1.4$~M$_\odot$. By doing this, we were able to match (at least
partially) the Roche lobe to the emission feature in the tomogram for
the cases with mass ratios ranging $0.1<q<0.4$ and therefore for
semiamplitudes of the secondary ranging between $190<{K_2}<230$~km
s$^{-1}$. Fig. 5 displays the predicted positions of the Roche Lobe of
the secondary star and the theoretical paths of the gas stream for an
averaged mass ratio of $q={{M_2}/{M_1}}=0.3$ and a semiamplitude of
the secondary of $K_2$=210 km s$^{-1}$. Note that the masses implied
for the secondary star of $0.1<{M_2}<0.6$~M$_\odot$ are consistent
with the range of masses obtained by applying the properties of a
stripped giant to the secondary of AC211 (see Homer et
al. 1998). Additionally, for a fixed inclination of 90 degrees we
found that for ${M_1}\geq3$~M$_\odot$ the Roche lobe did not match the
emission in the tomogram, which would suggest that the primary is a
neutron star. However, these conclusions are based on the assumption
that the S-wave arises from the secondary star: the fact that the
S-wave component can sometimes be strong during the eclipse ingress
(see upper panel Fig. 3), at the time when the X-ray irradiated face
of the secondary (the expected main contributor to the S-wave
component flux) points away from the observer, implies that the
S-wave may not always originate on the secondary star and that the
dynamical constraints obtained above should be treated with
caution.

In addition, another result from the Doppler map is the absence of
hotspot (the stream/disc impact region) expected at
${V_X}<0$,${V_Y}>0$ and the lack of signs of reimpact of the gas
stream in the lower quadrants (${V_Y}<0$). The latter is an unexpected
result for AC211: as stated previously, one might hope that the broad
dip centred at phase 0.5 in the optical light curve could be due to
occultation of the ADC/accretion disc by a bulge at the rim of the
accretion disc caused by the reimpact of outflowing gas (see Sec
1). This map does not appear to support this interpretation unless the
hotspot and the region of reimpact are fainter or comparable in
brightness to the accretion disc at He{\sc ii}~$\lambda4686$.

Nonetheless, it is tempting to associate the unusual 
spectroscopic properties of AC211 with its origin via tidal
capture. Immediately after formation it is likely that the secondary
was misaligned with respect to the orbital angular momentum
vector. If the secondary spin and the orbital angular momentum
vectors are not yet aligned, the gas stream will leave the inner
Lagrangian point of the donor star with an additional kinematical
contribution in the direction perpendicular to the orbital plane (see
e.g. Lanzafame, Belvedere \& Molteni 1994) and travel a distance above
the orbital plane considerably larger than the vertical height of the
accretion disc. Hence, the gas stream will be visible (generating the
enhanced emission in the Doppler map), travel without interaction with
the outer edge of the accretion disc, and fall back to the disc,
hitting its inner side. Thus the impact of the gas stream will create
a bulge which may cause the dip in the light curves centred at phase
0.5 with no observable bright spot since the hot regions are not
dominant at He{\sc ii}~$\lambda4686$ and/or they are obscured by the
outer disc. In this scenario, the variations observed in the intensity
of the S-wave and the long-term changes in the He{\sc ii} line
profiles are due to long term changes in the accretion geometry
attributable in turn to a precessing secondary star. Indeed, there is
some observational evidence for binary stars with one component
rotating asynchronously and with a misaligned spin (Gl\c ebocki \&
Stawikowski 1997).

Although this is an interesting model, the current theories of tidal
evolution of close binary systems predict that the alignment between
the orbital axis and the rotational axis of a secondary with a
convective envelope should occur earlier than or at a similar time to
the circularization of the orbit (see Hut 1981, Eggleton \&
Kiseleva-Eggleton 2001). In particular, the tidal circularization
time-scale for AC211 is estimated to be a few hundred years (Eggleton
2002). Therefore it is highly improbable that we are observing a
misaligned secondary star unless the system has been formed recently
or the secondary underfills its Roche lobe (both unlikely scenarios).

\section{Summary and conclusions}

In this paper we have presented time-resolved optical spectroscopy of
AC211/X2127+119. Our Doppler tomogram of the He{\sc ii}~$\lambda4686$
line shows enhanced emission at ${V_X}=30$~km s$^{-1}$, ${V_Y}=160$~km
s$^{-1}$ superimposed on emission arising from the accretion disc. The
phasing (if not the intensity modulation) of this emission suggests an
origin on the irradiated face of the secondary. This in turn implies
that AC211 is a high mass ratio system ($0.1<q<0.4$). Furthermore, the
Doppler map suggests a neutron star primary and a low mass secondary
star of mass $0.1<{M_2}<0.6$~M$_\odot$. We observe that the He{\sc
ii}~$\lambda4686$ shows considerable variation in its line profile on
a time-scale of order a few days clearly unrelated to any orbital
modulation. These have also been noted by van Zyl et
al. (2002). Further observations are required in order to understand
these changes in the emitting gas structure. Although a misaligned
secondary (as might be expected immediately after tidal capture) may
provide a natural explanation of some of the phenomena observed here,
it is excluded on the basis of the short circularization time-scale of
the orbit.

\section{ACKNOWLEDGEMENTS}

Use of {\sl MOLLY}, {\sl DOPPLER} and {\sl TRAILER} developed largely
by T. R. Marsh is acknowledged. We are grateful to P. Garnavich and
R. P. Kirshner for generously obtaining some of the data used in this
paper. We thank Craig Foltz for helpful comments during the data
reduction, Lee Homer for providing the AC211 linear ephemeris,
A. Recio-Blanco for confirming the presence of Ti{\sc
ii}~$\lambda4468.5$ in Horizontal-Branch stars in M15 and P. Eggleton
for his remarks about the possible tidal evolution of AC211. We also
thank the anonymous referee for useful comments.

\clearpage
\begin{table}
\caption{Journal of Observations.}
\label{tablelog}
\begin{center}
\begin{tabular}{lccccccc}
Date & No. spectra & HJD start     & HJD end 	& $\lambda$ range & Spatial Scale~$^{a}$ & Slit Width	\\ 
(UT) &             & (+2,450,000.) & (+2,450,000.)  & (\AA)           & (arcsec/pix)	& (arcsec)	\\
\\
10/08/96        &	   9 	   & 305.7836 & 305.9164 &   4240-5190     &  0.3  &	1  \\
11/08/96 	&	   6 	   & 306.8062 & 306.9930 &       "         &   "   &	"  \\
12/08/96 	&	  21 	   & 307.7018 & 307.9941 &   4335-5282     &   "   &	"  \\
13/08/96 	&	   5 	   & 308.7966 & 308.8337 &   4253-5190     &  0.6  &	"  \\
14/08/96	&	   9 	   & 309.7753 & 309.8507 &       "         &  0.6  &	"  \\
31/08/96 	&	  17   	   & 326.7333 & 326.9509 &   4236-5179     &   "   & 1.25  \\     
01/09/96 	&	   4       & 327.6787 &  327.8672 &  4388-5330     &   "   &    "  \\
02/09/96        &	   6       & 328.6290 &  328.9550 &      "  	   &   "   &    "  \\ 
\end{tabular}
\begin{tabular}{l} 
\\
$a$: the nominal unbinned scale is 0.3 arcsec/pix.
\end{tabular}
\end{center}
\end{table}

\clearpage
\begin{figure}
\epsfig{width=5in,file=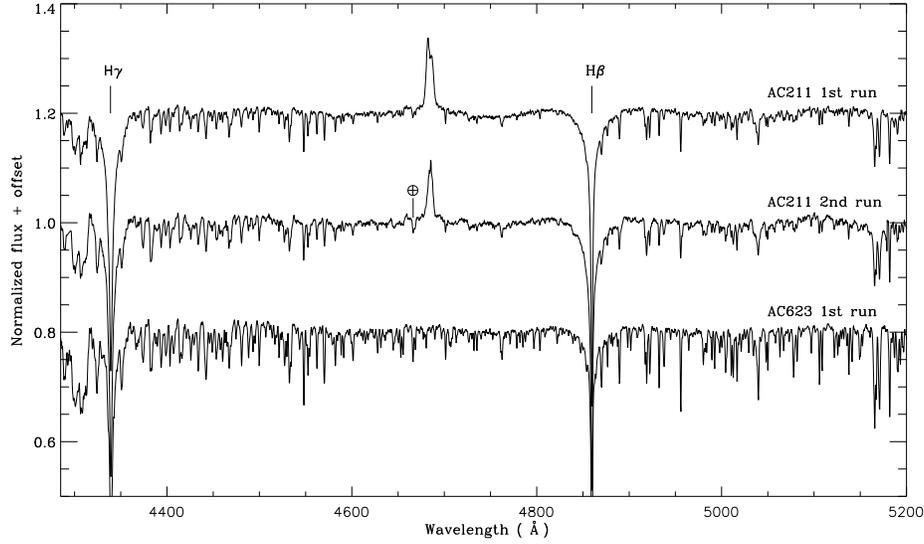}
\caption[mean.eps]{Normalized mean spectra of AC211 for the first
(top) and second (middle) runs, together with the mean spectrum of the
field star AC623 (bottom). The AC211 spectrum
shows the clear presence of He{\sc ii}~$\lambda4686$ in emission and
weak absorption features. The latter are totally due to background
contamination from neighbouring stars. $\oplus$ marks an instrumental
feature poorly removed by the flat field correction for the second
run.\label{fig2}}
\end{figure}  

\clearpage
\begin{figure}
\epsfig{width=5in,file=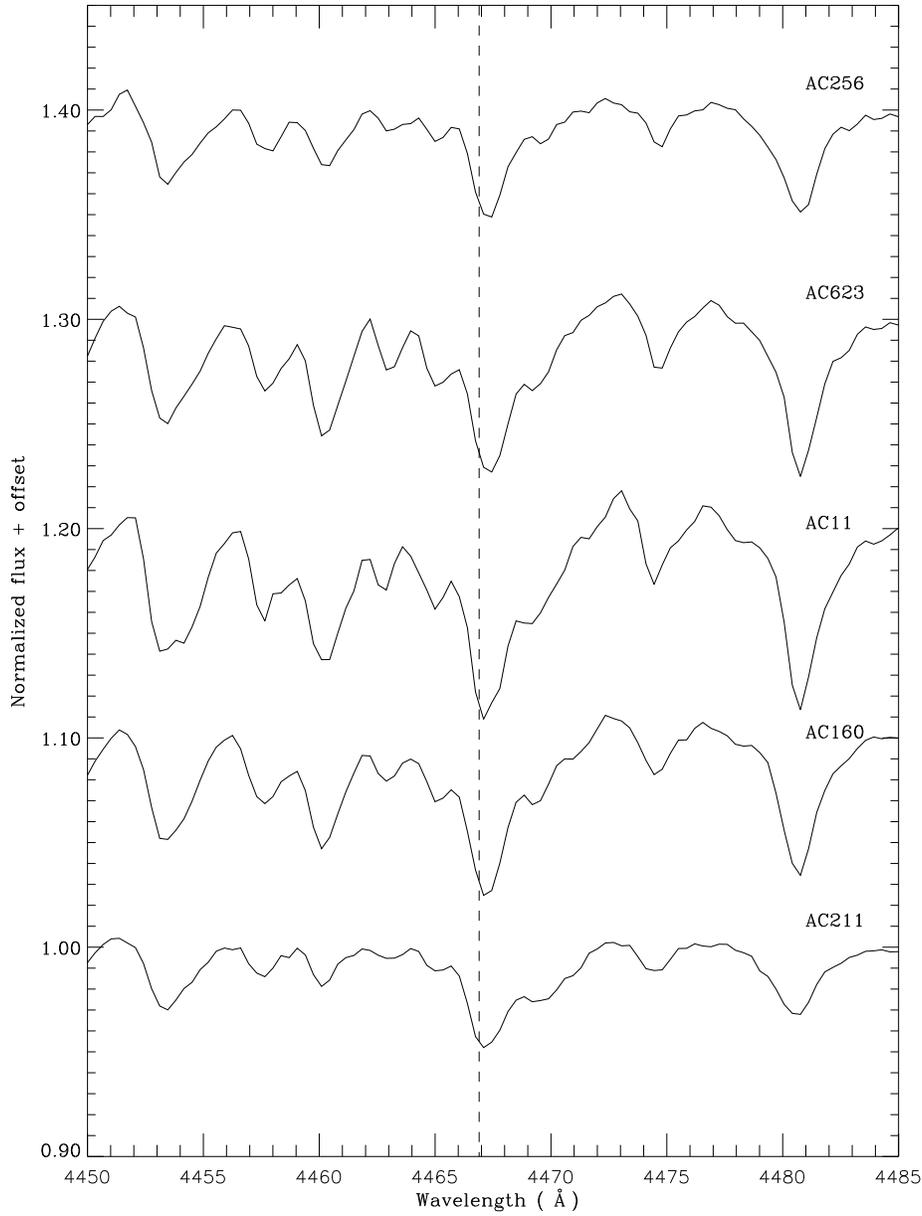}
\caption[evolutionnew.eps]{Enlargement at $\lambda4471$~for the mean
spectrum of AC211 and its neighbours. The dashed line marks the
expected position of the Ti{\sc ii}~$\lambda4468.5$~absorption line at
the globular cluster rest frame. \label{fig3}}
\end{figure}

\clearpage
\begin{figure}
\epsfig{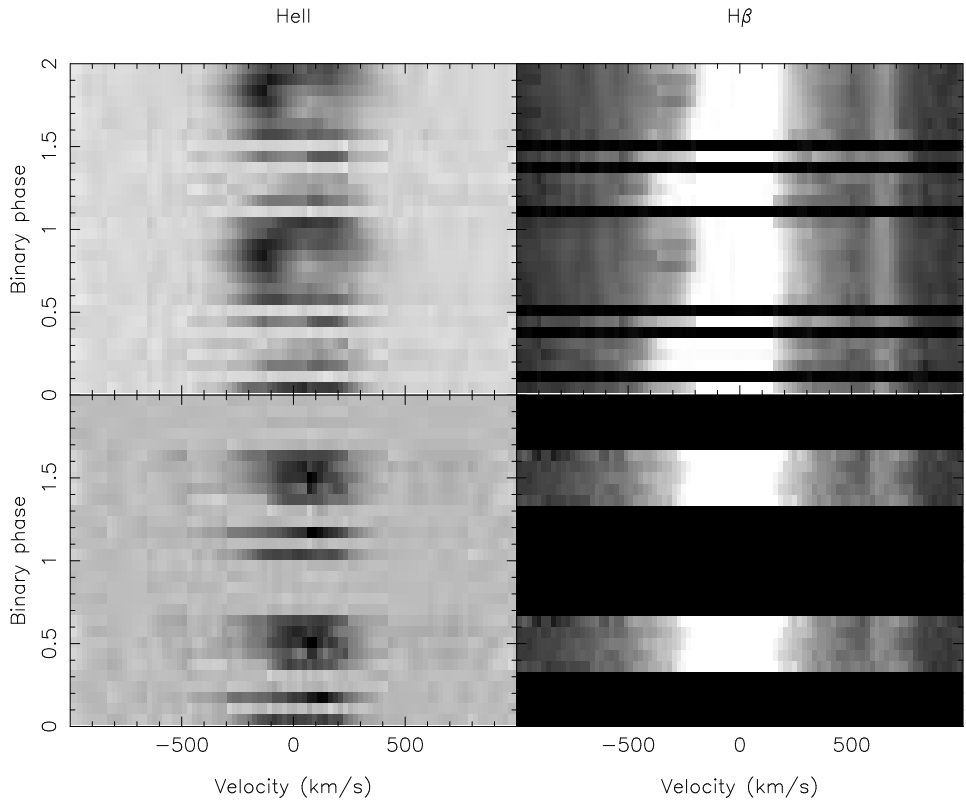}
\caption[trail.eps]{Trailed spectrograms of He{\sc ii}~$\lambda4686$~
and H$\beta$ for the first (upper panels) and second (lower panels)
run. For the sake of clarity, the same cycle has been plotted
twice. Empty strips represent gaps in the phase coverage. Zero
velocity corresponds to the laboratory wavelength. The grey scale
shows emission lines in black and absorption lines in white. The scale
in the H$\beta$ trailed spectrogram was chosen in order to outline the
blue wing absorption component of the line. Note that for the second
run, we removed the data acquired during the night of September 2
because the H$\beta$ profile fell on a bad column in the CCD
chip. \label{fig3}}
\end{figure}

\clearpage
\begin{figure}
\epsfig{,width=5in,file=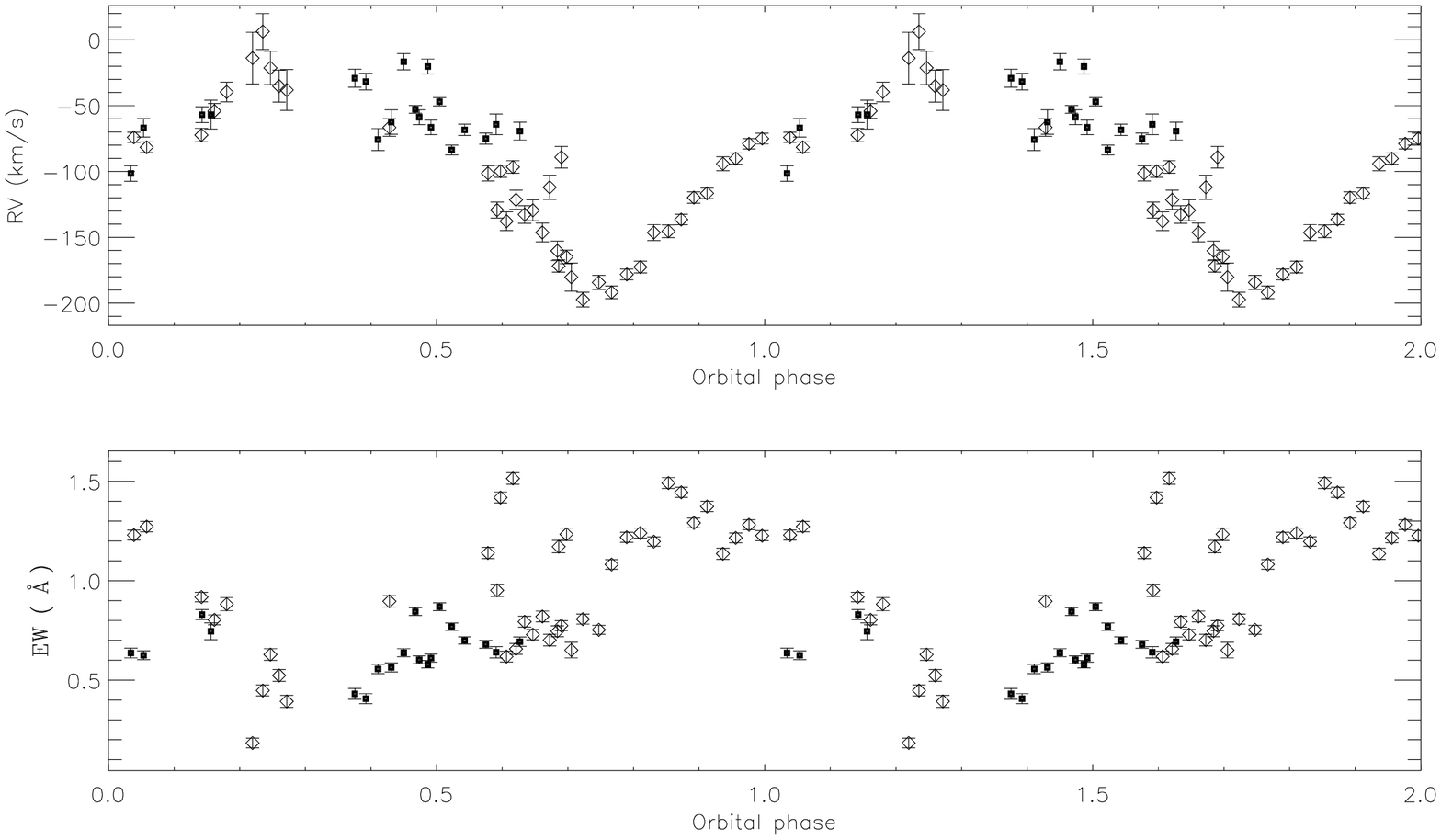}
\caption[rv.eps]{Radial velocities (upper panel) and EW (lower panel)
of He{\sc ii}~$\lambda4686$~folded on the ephemeris of Sec. 2. Diamonds
and solid squares correspond to the first and second run
respectively. The same cycle has been plotted twice.\label{fig3}}
\end{figure}

\clearpage    
\begin{figure}
\epsfig{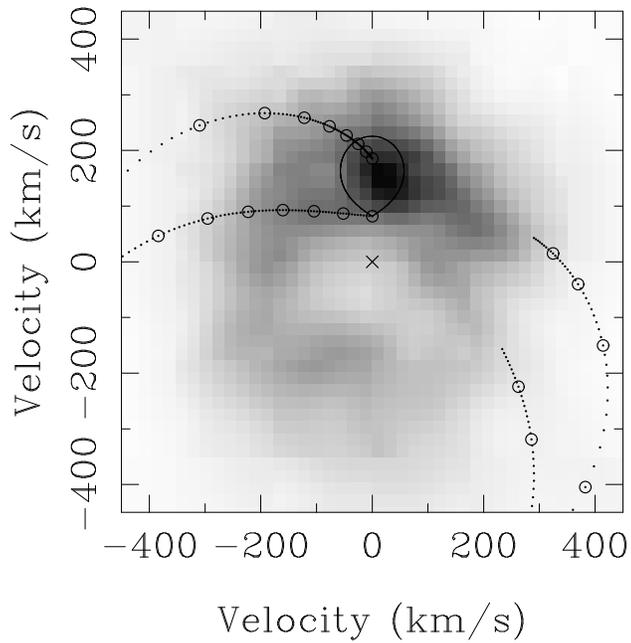}
\caption[map.eps]{Doppler map of He{\sc ii}~$\lambda4686$ computed by
using the whole 40 spectra from the first run. The Roche lobe of the secondary star, the predicted velocities of the gas stream (lower curve) and of
the disc along the gas stream (upper curve) are plotted for $K_2=210$~km s$^{-1}$
and $q={{M_2}/{M_1}}=0.3$. Distances in units of 0.1$R_{\rm L1}$ are marked along both
 curves with open circles. The centre of mass of the system is denoted by a cross. \label{fig6}}
\end{figure}


\begin{thebibliography}{}

\bibitem[]{}Auri\'ere M., Cordoni J.-P., 1981, A\&AS, 46, 347

\bibitem[]{}Auri\'ere M., le F\'evre O., Terzan A., 1984, A\&A, 138, 415

\bibitem[]{}Auri\'ere M., Maucherat A., Cordoni J.-P, Fort B, Picat
J. P., 1986, A\&A, 158, 158

\bibitem[]{}Bailyn C. D., Garcia M., Grindlay J. E., 1989, ApJ, 344, 786

\bibitem[]{}Behr B. B., Cohen J. G., McCarthy J. K., 2000, ApJ, 531, 37

\bibitem[]{}Callanan P. J., Naylor T., Charles P. A., 1990, in Mauche C. W.,
eds, Accretion-Powered Compact Binaries. Cambridge University Press, Cambridge, p. 51

\bibitem[]{}Callanan P. J., Drake J. J., Fruscione A., 1999, ApJ, 521, L125

\bibitem[]{}Charles P. A., Jones D. C., Naylor T., 1986, Nat., 323, 417

\bibitem[]{}Clark G. W., 1975, ApJ, 199, L143

\bibitem[]{}Cohen J. G., 1979, ApJ, 231, 751

\bibitem[]{}Deutsch E. W., Margon B., Anderson F. S., 2000, ApJ, 530, L21

\bibitem[]{}Dotani T., Inoue H., Murakami T., Nagase F., Tanaka Y., 1990, Nat., 347, 534

\bibitem[]{}Downes R. A., Anderson S. F., Margon B., 1996, PASP, 108, 688

\bibitem[]{}Eggleton P. P., Kiseleva-Eggleton L., 2001, ApJ, 562, 1012

\bibitem[]{}Eggleton P., 2002, private communication

\bibitem[]{}Elson R., Hut P., Inagaki S., 1987, ARAA, 25, 565

\bibitem[]{}Fabian A. C., Pringle J. E., Rees M. J., 1975, MNRAS, 172, 15P

\bibitem[]{}Fabian A. C., Guilbert P. W., Callanan P. J., 1987, MNRAS, 225, 29P

\bibitem[]{}Gl\c ebocki R., Stawikowski A., 1997, A\&A, 328, 579

\bibitem[]{}Gebhardt K., Pryr C., Williams T. B., Hesser J. E., Stetson P. B., 1997, ApJ, 113, 1026

\bibitem[]{}Giacconi R., Murray S., Gursky H., Kellogg E., Schreier
E., Matilsky T., Koch D., Tananbaum H., 1974, ApJS, 27, 37

\bibitem[]{}Grindlay J. E., 1993, The Globular Cluster-Galaxy Connection, ed. Graeme H. S. and Brodie J. P., ASP, 48, 156

\bibitem[]{}Hertz P., Grindlay J. E., 1983, ApJ, 275, 105

\bibitem[]{}Hertz P., 1987, ApJ, 315, L119

\bibitem[]{}Homer L., 2002, private communication

\bibitem[]{}Homer L., Charles. P. A., 1998, New Astron., 3, 435

\bibitem[]{}Hut P., 1981, A\&A, 99, 126

\bibitem[]{}Hut P., McMillan S. L. W., Goodman J., Mateo M., Phinney E. S., Pryor C., Richer H. B., Verbunt F., Weinberg M., 1992, PASP, 104, 981

\bibitem[]{}Ilovaisky S. A., Auri\'ere M., Kock-Miramond L., Chevalier C., Cordoni J. P., 1987, A\&A, 179, L1 

\bibitem[]{}Ilovaisky S. A., 1989, Proc of the $23^{rd}$ ESLAB Symposium on Two Topics in X-ray Astronomy, Volume 1, p. 145

\bibitem[]{}Ilovaisky S. A., Auri\`ere M., Koch-Miramond L., Chevalier C., Cordoni J. P., Crowe R. A., 1993, A\&A, 270, 139

\bibitem[]{}Ioannou Z., Naylor T., Smale A. P., Charles P. A., Mukai K., 2002a, A\&A, 382, 130

\bibitem[]{}Ioannou Z., Naylor T., van Zyl L., Charles P. A., Smale
A. P., Mukai K., 2002b, in G\"ansicke B. T., Beuermann K. \& Reinsch K.,
eds, ASP Conf. Ser. Vol. 261, The Physics of Cataclysmic Variables and
Related Objects. Astron. Soc. Pac., San Francisco, p. 351

\bibitem[]{}Katz J. I., 1975, Nat., 253, 698

\bibitem[]{}Lanzafame G., Belvedere G., Molteni D., 1994, MNRAS, 267, 312

\bibitem[]{}Marsh, T. R., Horne K., 1988, MNRAS, 235, 269

\bibitem[]{}Marsh, T. R., Horne K., 1990, ApJ, 349, 593

\bibitem[]{}Massey P., Foltz C., 2000, PASP, 112, 566

\bibitem[]{}Naylor T., Charles P. A., Drew J. E., Hassall B. J. M., 1988, MNRAS, 233, 285

\bibitem[]{}Naylor T., Charles P. A., 1989, MNRAS, 236, 285

\bibitem[]{}Naylor T., Charles P. A., Hasall B. J. M., Raymond J. C., Nassiopoulos G., 1992, MNRAS, 255, 1

\bibitem[]{}Smale A. P., 2001, ApJ, 562, 957

\bibitem[]{}Sneden C., Johnson J., Kraft R. P., Smith G. H., Cowan J. J., Bolte M. S., 2000, ApJ, 536, 85


\bibitem[]{}van Paradijs J, Dotani T., Tanaka Y., Tsuru T., 1990, PASJ, 42, 633 

\bibitem[]{}van Zyl L., Naylor T., Charles P. A., Ioannou Z., to appear in MNRAS

\bibitem[]{}Verbunt F., Van den Heuvel E. P. J., 1995, in Lewin W. H. G., van Paradijs J. \& van den Heuvel E. P. J., eds, X-Ray Binaries. Cambridge University Press, Cambridge, p. 457

\bibitem[]{}White N. E., Holt S. S., 1982, ApJ, 257, 318

\bibitem[]{}White N. E., Angelini L., 2001, ApJ, 561, L101


\end{thebibliography}
\end{document}